\documentclass[twocolumn, reprint]{revtex4-2}
\usepackage{natbib}
\usepackage{graphicx}
\usepackage{amsmath}
\usepackage{amssymb}
\usepackage{gensymb}
\usepackage{color}
\usepackage{url}

\begin{document}
\title{Coherent vibrational dynamics in molecular bond breaking: methyl radical umbrella mode probed by femtosecond x-ray spectroscopy}
\author{Christian A. Schr\"oder${}^{*,1,2,\dagger}$, John H. Hack${}^{1,2,\dagger}$, Joshua L. Edwards${}^{1,2}$, Zhiyu Zhang${}^{1,2}$,J. Tyler Kenyon${}^{1,2}$, Qiyue Wang${}^{4}$, Han Wang${}^{4}$, Daniel M. Neumark${}^{1,2}$, Stephen R. Leone${}^{1,2,3}$\\
\small{${}^{1}$Department of Chemistry, University of California, Berkeley, CA 94720, USA.\\
${}^{2}$Chemical Sciences Division, Lawrence Berkeley National Laboratory, Berkeley, CA 94720, USA.\\${}^{3}$Department of Physics, University of California, Berkeley, CA 94720, USA\\${}^{4}$State Key Laboratory of Quantum Functional Materials, School of Physical Science and Technology,\\ShanghaiTech University, Shanghai 201210, China\\${}^{*}$Contact author: \url{caschroeder@lbl.gov}},${}^{\dagger}$These authors contributed equally.}

\begin{abstract}
We report on the observation of coherent molecular vibrations launched by the breaking of a molecular bond. The methyl radical, which is produced by $267\,\mathrm{nm}$ photodissociation of methyl iodide, is excited to high levels in its $\nu_2$ ``umbrella" vibrational mode by the dissociation. The ensuing coherent vibrational dynamics  are observed by measuring ultrafast time-dependent changes in the x-ray transition energy from the C$1s$ to the singly-occupied valence orbital. Due to symmetry, the real space vibrational motion appears predominantly in the x-ray energy shift at the difference frequencies of the $\nu_2$ progression, although the fundamental frequencies of the $\nu_2$ mode are also observed. By constructing a fully quantum-mechanical model of the dynamics the coherent superposition is rigorously characterized and the real-space motion of the radicals is reconstructed. The retrieved trajectories are dominated by pronounced quantum beating governed by the high degree of coherent excitation and the strong negative anharmonicity of the $\nu_2$ mode.
\end{abstract}

\maketitle

\section{Introduction}
Impulsive excitations regularly induce coherent vibrational dynamics in molecules. While such an excitation is usually delivered to the system under study artificially e.g. with a short laser pulse \cite{chang2022conical, ou2024attosecond, rupprecht2023resolving, barreau2023core}, it was recently shown that the rapid change in geometry intrinsic to Jahn-Teller distortion can also produce the required ``kick" to the system \cite{ridente2023femtosecond}, leading to coherent dynamics. Whether the breaking of a chemical bond -- another fundamental process of chemical transformations -- is also able to impart coherent vibrational excitation into the resulting fragments is still an open question. In this letter we address coherence in bond breaking by reporting on the observation of coherent activation of the $\nu_2$ ``umbrella" vibration of the methyl radical produced by the $267\,\mathrm{nm}$ photodissociation of methyl iodide, probed by ultrafast x-ray spectroscopy (cf. \cite{attar2017femtosecond, bhattacherjee2018ultrafast, ridente2023femtosecond}). Previous ultrafast studies have approached the process from the perspective of the iodine atom, focusing on resolving the transition state region \cite{attar2015direct} or the conical intersection \cite{chang2021mapping, chang2022conical}. Selectively probing the methyl radical and its dynamics, this experiment provides a complement to these previous studies that allows vibrational coherence to be identified.
\begin{figure*}
	\includegraphics{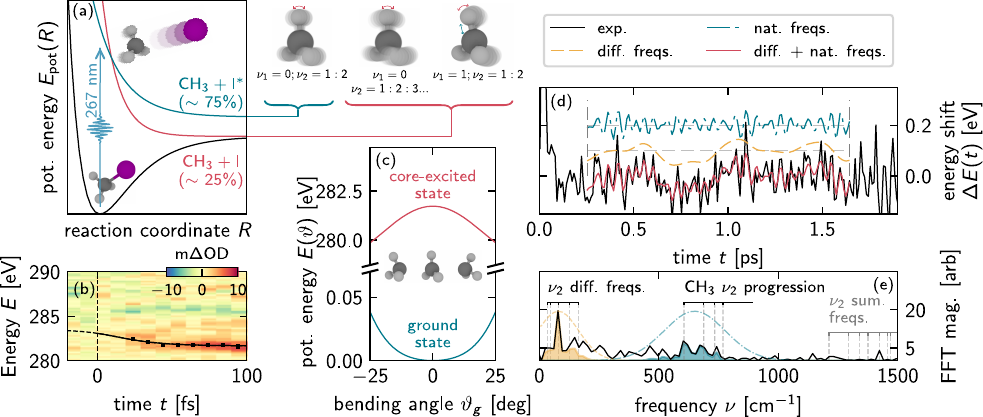}
	\caption{(a) Dissociation of methyl iodide probed with femtosecond x-ray spectroscopy. Excitation via $267\,\mathrm{nm}$ radiation will promote the molecule onto its ${}^{3}Q_0$ surface (blue curve) from where it will evolve towards a conical intersection with the ${}^{1}Q_{1}$ surface (red). The two ensembles of radicals that are produced from the dissociation are shown as ball-and-stick models with their vibrational excitation indicated. (b) The first $100\,\mathrm{fs}$ of the transient absorption signal shows the radical peak initially appearing at $282.9\,\mathrm{eV}$ and moving to its dissociation limit value of $281.7\,\mathrm{eV}$. Black dots are the peak position. We determine a time constant of $(20.7\pm 5.4)\,\mathrm{fs}$ by fitting an exponential decay convolved with a Gaussian function of $36\,\mathrm{fs}$ width for the instrument response (black curve). The depletion of the ground state absorption near $286\,\mathrm{eV}$ is also visible. (c) Potential energy surfaces for the methyl radical along the $\nu_2$ umbrella vibrational motion for the ground state and core-excited state as a function of the bending angle $\vartheta_g$ modeled here. (d) Energetic position of the radical peak as a function of pump-probe delay time (black) decomposed into the difference frequency component due to radicals excited only along the umbrella mode (yellow dashed) and the fundamental frequencies (blue dash-dotted), which become visible due to symmetry breaking (see text). Beginning and end of the filtered data are cropped as these contain the time-domain image of the filter windows. (e) Frequency domain picture of what is shown in panel (d) with the filter windows indicated. A clear peak is seen in the region where the difference frequencies are expected (yellow, dashed), as well as a series of peaks that increase in amplitude again and closely resemble the fundamental $\nu_2$ vibrational progression (blue, dash-dotted). Other features in the low frequency region are due to noise and not associated with any significant signal.}
	\label{img:main}
\end{figure*}
\section{Experimental}
The experimental apparatus is similar to that reported in \cite{barreau2020efficient}, modified to initiate dynamics with ultraviolet (UV) pump-pulses. Briefly, we split off part of the incoming $35\,\mathrm{fs}$ pulses centered at $800\,\mathrm{nm}$ in the pump-line and frequency double  it to obtain $40\,\mathrm{\mu J}$ of $400\,\mathrm{nm}$ light in a $200\,\mathrm{\mu m}$ thick $\beta$-barium borate (BBO) crystal. This is then recombined with $750\,\mathrm{\mu J}$ of $800\,\mathrm{nm}$ light in another BBO crystal ($50\,\mathrm{\mu m}$ thick) in vacuum to generate UV pulses of up to $20\,\mathrm{\mu J}$ pulse energy with a spectral width of $3\,\mathrm{nm}$, centered around $267\,\mathrm{nm}$. Measured via a cross-correlation in argon gas the apparatus has a temporal instrument response of $(36\pm 6)\,\mathrm{fs}$ and its spectral resolution is $\sim 300\,\mathrm{meV}$ near $244\,\mathrm{eV}$ as measured via the static absorption of argon gas at its $L_{2, 3}$ edge. Methyl iodide is obtained commercially at $>98\%$ purity and is used without further purification. The liquid readily forms sufficient vapor pressure for the experiment at room temperature and is introduced into a $4\,\mathrm{mm}$ long gas cell with $200\,\mathrm{\mu m}$ openings, which forms the interaction region. The UV pump pulse and soft x-ray probe pulse cross in the gas cell at a small angle ($\sim 1\degree$). An iris is positioned after the gas cell and set to block the UV light while transmitting the soft x-ray beam into the spectrometer assembly. Sequences of spectra are collected as a function of delay time, both with the pump beam irradiating the sample and with it blocked. The change in absorption due to the presence of the pump pulse is calculated as $\Delta\mathrm{OD} = - \log_{10}\left(S_\mathrm{on} / S_\mathrm{off}\right)$, where $S_\mathrm{on/off}$ denotes spectra taken with the pump beam on and off respectively. We chose time steps of $\Delta t = 10\,\mathrm{fs}$ and a delay time range of approximately $2\,\mathrm{ps}$. Individual delay scans recorded over the course of multiple days are aligned in time and averaged.
\section{Coherent vibrations in methyl iodide photodissociation}
\label{sec:results}
In order to launch a coherent vibration, the excess kinetic energy in the dissociation has to be transferred to the emerging methyl radicals on a timescale shorter than a vibrational period. Photodissociation of methyl iodide in its $A$-band near $270\,\mathrm{nm}$ occurs either on the strongly repulsive ${}^{3}Q_0$ surface ($75\%$ yield), which is initially populated by the UV pump pulse, or via a conical intersection to access the ${}^{1}Q_{1}$ surface ($25\%$ yield) that is approached within $(15\pm4)\,\mathrm{fs}$ \cite{chang2022conical} (see fig. \ref{img:main}(a)). Radicals originating from the ${}^{3}Q_0$ surface are vibrationally excited up to $\nu_2 = 2$ and will leave behind electronically excited iodine atoms in the ${}^{2}P_\frac{1}{2}$ state (denoted as $\mathrm{I^*}$ channel), while for those that emerge via the ${}^{1}Q_{1}$ surface the energy partitioning forms the iodine atom's ground state (${}^{2}P_\frac{3}{2}$, denoted $\mathrm{I}$ channel), and the radical is excited up to $\nu_2 = 4$ \cite{eppink1999energy, li2005high, aguirre2005photoionization, cheng2011vibrationally} (see fig. \ref{img:main}(a)).
We can assign a time constant to the dissociation process by tracking the initial energy shift of the characteristic C$1s\to$SOMO (singly-occupied molecular orbital) x-ray absorption peak of the methyl radical, and find that it reaches its dissociation-limit value of $281.7\,\mathrm{eV}$ within $\tau_\mathrm{d} = (20.7 \pm 5.4)\,\mathrm{fs}$ (see fig. \ref{img:main}(b)). Together with the temporal instrument response of $\tau_\mathrm{IRF} = (36.0 \pm 6.4)\,\mathrm{fs}$ this confines the duration of the impulse delivered to the methyl radicals in the experiment to an upper limit of $(41.5 \pm 6.2)\,\mathrm{fs}$, or $\sim 75\%$ of the oscillation period of the $\nu_2 = 1\to 0$ transition ($55\,\mathrm{fs}$). This suggests that coherent vibrational dynamics can in principle be induced via bond breaking in the experiment. Ultrafast extreme-ultraviolet (XUV) or soft x-ray spectroscopy resolves coherent molecular vibrations by measuring time-dependent changes in the x-ray transition energy as the molecule moves along a vibrational coordinate on the respective potential energy surfaces (cf. \cite{barreau2023core, rupprecht2023resolving, ou2024attosecond, chang2022conical, chang2021mapping}). If the change in transition energy $E_\mathrm{cx}(q)$ as a function of the reaction $q$ coordinate can be approximated well by a linear function, then the energy shift will be directly proportional to the real space motion along the trajectory $\langle q(t)\rangle$. The methyl radical's umbrella vibration, however, is completely symmetric with respect to its reaction coordinate (the bending angle $\vartheta_g$). As such, the change in transition energy $E_\mathrm{cx}(\vartheta_g) \propto \vartheta_g^2 + \dots$ must be an even function of $\vartheta_g$. The trajectory $\langle\vartheta_g(t)\rangle$, which oscillates at frequencies $\Delta\omega_{1\to0}, \Delta\omega_{2\to1}, \Delta\omega_{3\to2}, \dots$ if a coherent superposition of multiple vibrational states is formed, will be mapped onto an energy shift that oscillates at the sums and differences of these frequencies. Unless the symmetry of the radical's motion or that of its mapping onto the x-ray energy is broken, the fundamental frequencies will be quenched in the energy shift's frequency content. The implication for the experiment is that one signature of coherent $\nu_2$ activation is slow oscillatory motion of the C$1s\to$SOMO energy with a period on the order of $\sim 400\,\mathrm{fs}$ ($80\,\mathrm{cm^{-1}}$ instead of $607\,\mathrm{cm^{-1}}, 688\,\mathrm{cm^{-1}}, \dots$ \cite{oliveira2015photoelectron}) due to the difference frequencies, superimposed with a rapid oscillation due to the sum frequencies with about $25\,\mathrm{fs}$ ($\sim 1300\,\mathrm{cm^{-1}}$) period. Similar arguments regarding symmetry and the detection of mixing products of vibrational frequencies have been brought forward by \citet{yatsuhashi2004ultrafast} and \citet{fuss2007ultrafast} (see also references therein), but in the context of ultrafast mass-spectroscopy. Figure \ref{img:main}(d) and (e) show the evolution of the radical peak's position as a function of time after photodissociation as measured in the experiment and its Fourier transform respectively. A broad feature with a strong peak near $80\,\mathrm{cm^{-1}}$ that lines up exactly with the difference frequency between the $\nu_2 = 2\to 1$ and $1\to 0$ transitions is clearly present (yellow shaded area in fig. \ref{img:main}(e)), showing unambiguously that vibrational coherence in the umbrella mode up to at least $\nu_2 = 2$ is induced and is observable through the difference frequency signal. The rapidly oscillating sum frequencies, which would be found near $1300\,\mathrm{cm^{-1}}$ and above, are not observed in the experiment due to insufficient temporal resolution. Quite interestingly, there is also a series of peaks that precisely match the $\nu_2$ umbrella progression at its fundamental frequency up to the $\nu_2 = 3\to 2$ transition (shaded in blue). The appearance of these peaks requires the symmetry assumed in the discussion above to be broken, which is elaborated upon further in \ref{sec:chstretch}. 
\begin{figure}
	\includegraphics{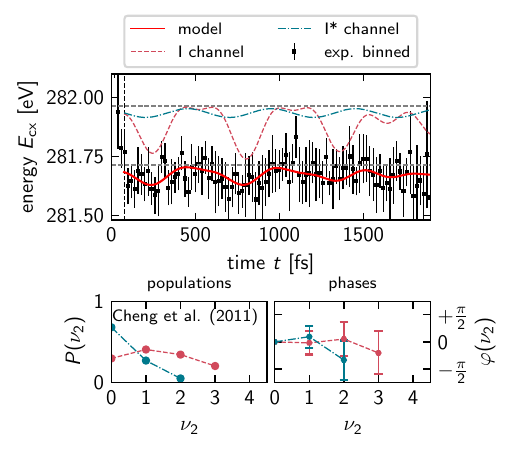}
	\caption{Application of the two-channel model to the experiment using the populations reported by \citet{cheng2011vibrationally}. The points in the top panel is the experimental data, where adjacent points have been binned for the fit. The bold red line is the fit of our model. In all panels, blue dash dotted and red dashed lines indicate results for the $\mathrm{I^{*}}$ and $\mathrm{I}$ channel, respectively. The individual energy shifts in the $\mathrm{I}$ and $\mathrm{I^*}$ channels are also shown in the top panel, with a shift applied for visual clarity. Their weighted sum $E_{\mathrm{exp.}}(t) \approx \sigma_\mathrm{I^*} E_\mathrm{cx}^\mathrm{I^*}(t) + (1 - \sigma_\mathrm{I^*})E_\mathrm{cx}^\mathrm{I}(t)$ yields the bold red line.}
	\label{img:het_fitting}
\end{figure}
\subsection{Two-channel model of coherent vibrational dynamics}
\label{sec:model}
In order to further corroborate that vibrational coherence is induced in the experiment and extract phase information that fully characterizes the quantum superpositions in which the umbrella mode is populated in both the $\mathrm{I}$ and $\mathrm{I^*}$ channel, we construct a model based on a quartic vibrational potential $V_\mathrm{gs}(\vartheta)$. The use of a quartic term was identified as an appropriate description of the umbrella mode in early studies of the methyl radical (e.g. \cite{hermann1982photofragmentation, hermann1982photofragment, yamada1981diode}), as it captures the pronounced negative anharmonicity of the $\nu_2$ mode well. Following \citet{ragni2016umbrella}, the umbrella motion is described via the hyperangle $\vartheta$ that captures the motion along the true $\nu_2$ bending angle $\vartheta_g$ and the small changes in the $\mathrm{C-H}$ bond length $r_\mathrm{CH}$ that accompany $\nu_2$ activation (see supplemental material and fig. \ref{img:real_space}(a, b)). For the potential here $V_\mathrm{gs}(\vartheta) = E_1^\mathrm{gs}\vartheta^2 + E_2^\mathrm{gs}\vartheta^4$, and the parameters $E_1^\mathrm{gs}$ and $E_2^\mathrm{gs}$ are optimized such that the transition energies of the potential match tabulated values \cite{oliveira2015photoelectron}. It is shown as the blue curve in fig. \ref{img:main}(c). A coherent superposition
\begin{equation}
	\psi(\vartheta, t) =
		\sum_{i=0}^{\nu_\mathrm{max}}
		\left|a_i\right|\phi_i(\vartheta)\mathrm{e}^{-i\omega_it}\mathrm{e}^{-i\varphi_i}
		\label{eq:superposition}
\end{equation}
of its eigenfunctions $\phi_i(\vartheta)$, where $a_i$ are the amplitudes, $\omega_i$ are the eigenenergies and $\varphi_i$ the individual phases then describes the microscopic dynamics of the methyl radicals in which only the umbrella mode is activated. The displacement of the angle $\vartheta - \vartheta_0$ is
\begin{align}
    &\nonumber\langle \vartheta(t) - \vartheta_0\rangle =\\
    &2 \sum_{i = 1}^{\nu_\mathrm{max}}\left|a_i\right|\left|a_{i - 1}\right| \mu_{i, i - 1} \cos\left(\Delta\omega_{i\to i-1}t + \Delta\varphi_{i, i-1}\right),
    \label{eq:position}
\end{align}
where $\mu_{i, j} = \left\langle\left.\phi_i\right.\right|\vartheta - \vartheta_0\left.\left|\phi_j\right.\right\rangle$, $\Delta\omega_{i\to j} = \omega_i - \omega_{j}$, $\Delta\varphi_{i, j} = \varphi_i - \varphi_{j}$ and $\vartheta_0 = \pi/2$ is the equilibrium angle in the hyperspherical coordinate system, which is dropped for notational clarity in the following. With the appropriate phases and amplitudes, evaluating the energy of the core-excited state $E_\mathrm{cx}\left(\langle\vartheta(t)\rangle\right) = V_\mathrm{ce}\left(\langle\vartheta(t)\rangle\right) - V_\mathrm{gs}\left(\langle\vartheta(t)\rangle\right)$ as a function of the trajectory of $\langle\vartheta(t)\rangle$ will describe the evolution of the energetic position of the radical peak in either the $\mathrm{I}$ or $\mathrm{I^*}$ channel. In the experiment, these signals overlap, but the overall energetic position of the C$1s\to$SOMO peak moves as $E_{\mathrm{exp.}}(t) \approx \sigma_\mathrm{I^*} E_\mathrm{cx}^\mathrm{I^*}(t) + (1 - \sigma_\mathrm{I^*})E_\mathrm{cx}^\mathrm{I}(t)$, where $\sigma_\mathrm{I^*} \approx 0.75$ is the branching ratio at the conical intersection \cite{eppink1998methyl, gardiner2015dynamics}. We make the simplifying assumption that $E_\mathrm{cx}(\vartheta) \approx E_0^\mathrm{cx} + E_1^\mathrm{cx}\vartheta^2$ is a good approximation for the dependency of the core-excited state energy on the bending angle, which is justified by the success of the model as seen below. The finite temporal resolution of the experiment is incorporated into the model by convolving $E_\mathrm{exp.}(t)$ with a Gaussian function of $36\,\mathrm{fs}$ width. Using vibrational population data reported in literature \cite{li2005high, cheng2011vibrationally, eppink1999energy, aguirre2005photoionization}, the phases $\varphi_i$ can then be determined via a nonlinear least-squares fit, which is restricted to pump-probe delays $\tau > 75\,\mathrm{fs}$, as the initial energy shift during the bond breaking is not accounted for in the model. In both channels the phase $\varphi_0$ is fixed to $\varphi_0=0$. Using the parameters $E_0^\mathrm{cx}$ and $E_1^\mathrm{cx}$ from the fit the core-excited potential energy surface $V_\mathrm{ce}\left(\vartheta_g\right)$ along the umbrella motion is determined. It is shown in fig. \ref{img:main}(c) as a red curve. A more detailed account of the fitting procedure is given in the supplemental material. The result of this fit is shown in fig. \ref{img:het_fitting} using the population data of ref. \cite{cheng2011vibrationally}, and we find no significant differences using population data from other studies (refs. \cite{li2005high, aguirre2005photoionization, eppink1998methyl}, see supplemental material). Despite its simplicity the model captures the time-dependence of the radical's x-ray absorption peak observed in the experiment very well and over the entire range of time delays. This confirms the model is an appropriate description of the microscopic dynamics for $t>75\,\mathrm{fs}$, further strengthening the claim that coherent vibrational dynamics are induced via the dissociation of the molecular bond.
\begin{figure}
	\includegraphics{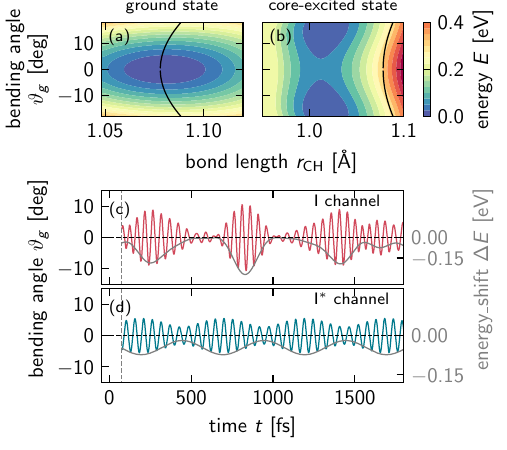}
	\caption{(a, b) Ground-state and core-excited state potential energy surfaces of \citet{ekstroem2008umbrella}. The path along which the real-space motion takes place is shown as black dashed lines. (c, d) Reconstructed real-space motion of the methyl radicals with pure $\nu_2$ activation along the angular coordinate $\vartheta_g$. The radial coordinate $r_\mathrm{CH}$ follows from the black dashed lines in panels (a) and (b). The energy shift (see also fig. \ref{img:het_fitting}) closely follows the envelope of the vibrational trajectory. The complex nature of the trajectory in the $\mathrm{I}$ channel is due to the pronounced beating resulting from the high degree of excitation. To a lesser, yet notable extent such beating is also present in the $\mathrm{I^{*}}$ channel trajectory.}
	\label{img:real_space}
\end{figure}
As now the phases $\varphi_i$, the amplitudes $a_i$ and the frequencies $\omega_i$ are determined (lower panels in fig. \ref{img:het_fitting}), the vibrational wavefunction (eq. \ref{eq:superposition}) is characterized in its entirety for $t > 75\,\mathrm{fs}$ in both dissociation channels. Therefore the real-space motion of the radicals can be reconstructed.

We evaluate eq. \ref{eq:position} and obtain the oscillatory trajectories shown in fig. \ref{img:real_space}(d) and (e). The signal in the $\mathrm{I}$ channel is dominated by pronounced beating over the entire range of delay times, leading to large excursions of the bending angle near $250\,\mathrm{fs}$, $830\,\mathrm{fs}$ and $1.1\,\mathrm{ps}$. While such beating is present also in the $\mathrm{I^*}$ channel, it is to a much lesser degree and the angular displacement is overall smaller, in accord with the lower degree of vibrational excitation. As we have approximated $E_\mathrm{cx}(\vartheta) \approx E_0^\mathrm{cx} + E_1^\mathrm{cx}\vartheta^2$, we can now compare our value $E_1^\mathrm{cx} = (-9.0 \pm 2.2)\,\mathrm{eV/rad^2}$ to quadratic expansions of energy curves calculated from first principles along the trajectory $\left(\langle\vartheta_g(t)\rangle,\langle r_\mathrm{CH}(t)\rangle\right)$. For the surfaces calculated by \citet{ekstroem2008umbrella} we find $E_1^\mathrm{cx} = (-2.89 \pm 0.01)\,\mathrm{eV/rad^2}$, which is significantly smaller in magnitude than the value determined from the fit. Likewise, we find $E_1^\mathrm{cx} = (-3.2 \pm 0.1)\,\mathrm{eV/rad^2}$ from a calculation of our own (see supplemental material), which compares well to the value determined for the surfaces of \citet{ekstroem2008umbrella}. For all cited population data though, we find $E_1^\mathrm{cx}$ to have a negative sign from the fit, indicating that the core-excited state potential energy surface has a double-well structure as it was found in \cite{ekstroem2008umbrella}. The discrepancies between the value of $E_1^\mathrm{cx}$ determined from the model fit and those determined from calculations may be due to how the overall energy shift $E_{\mathrm{exp.}}(t)$ is sampled in the experiment; the signals from the $\mathrm{I}$ and $\mathrm{I^*}$ channel overlap energetically and cannot be accurately disentangled, the expression $E_{\mathrm{exp.}}(t) \approx \sigma_\mathrm{I^*} E_\mathrm{cx}^\mathrm{I^*}(t) + (1 - \sigma_\mathrm{I^*})E_\mathrm{cx}^\mathrm{I}(t)$ we use for the peak position is only approximate. Despite discrepancies between the calculations and $E_1^\mathrm{cx}$ determined from the experiment using our model, it captures the essential aspects of the coherent dynamics of the methyl radical, i.e. the umbrella mode's pronounced anharmonicity, its symmetry and the resulting mapping of the real space coherent vibrational motion onto a slow oscillation of the observable energy shift.
\subsection{Symmetry breaking and observation of fundamental frequencies}
\label{sec:chstretch}
\begin{figure}
	\includegraphics{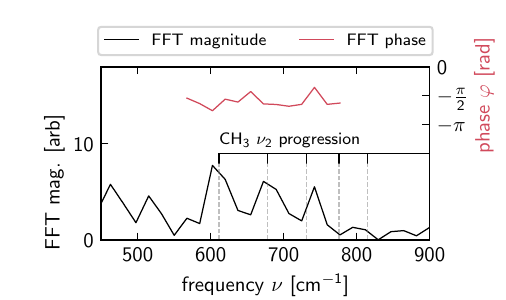}
	\caption{Direct observation of the $\nu_2$ vibrational progression mediated by symmetry breaking. The phase of this oscillation as determined via the Fourier transform (red) is essentially flat across this region. A slight detuning of the $\nu_2$ vibration, due to the concurrent activity in the $\nu_1$ mode may be the reason for the third peak near $750\,\mathrm{cm^{-1}}$ to appear at a slightly higher frequency (see text).}
	\label{img:umbrella_v1_phases}
\end{figure}
In the $(r_\mathrm{CH}, \vartheta_g)$-plane the methyl radical's umbrella mode moves along a trajectory of high symmetry (black curves in fig. \ref{img:real_space}(a, b)) leading to the difference frequency signal discussed above, with expected quenching of signal near the fundamental frequencies. The peaks in the $500-800\,\mathrm{cm^{-1}}$ region of the Fourier transform (fig. \ref{img:main}e, fig. \ref{img:umbrella_v1_phases}), however indicate the presence of a mechanism that breaks either the symmetry of the $\nu_2$ motion itself, or that of its mapping to the measured x-ray energy in either one or both of the dissociation channels. Symmetry breaking of the motion itself may be facilitated by the simultaneous activation of another vibrational mode. The $\mathrm{I}$-channel produces a small fraction ($\sim 12\%$ of its yield, see fig. \ref{img:main}a) of radicals in which the symmetric $\mathrm{CH}$-stretch $\nu_1$ mode is also activated \cite{li2005high} (albeit possibly not coherently in this experiment due to its high frequency and the time scale of bond breaking). If the resulting composite motion distorts the ideal quartic potential of the $\nu_2$ mode, it would break the symmetry and could account for some of the observed fundamental vibrational progression, possibly with a slight detuning. The peak near $750\,\mathrm{cm}$ is in fact slightly offset from the frequency determined by \citet{oliveira2015photoelectron}. However, the comparably large magnitude of these peaks despite the small fraction of radicals with $\nu_1$ activation is not accounted for. Another possible mechanism of transient symmetry breaking has been identified by \citet{kyriaki2010photodissociation}. Briefly, the transition probability of the wave packet from the electronic ground-state to the core-excited state may -- depending on the launched wave packet phase -- be favorable for bending angles near one of the two minima of the excited state surface (fig. \ref{img:real_space}b), if a small tunneling splitting is present, effectively breaking the symmetry on the timescale of the tunneling time. The splitting and associated tunneling time for the core-excited state of the methyl radical is $13.8\,\mathrm{meV}$, or $149\,\mathrm{fs}$, respectively \cite{ekstroem2008umbrella}, i.e. roughly three times the period of the fundamental $\nu_2$ vibration. Figure \ref{img:umbrella_v1_phases} shows the $500-800\,\mathrm{cm^{-1}}$ region of the x-ray energy shift's Fourier transform in more detail. Isolating the peaks with a Gaussian filter (blue curve in fig. \ref{img:main}(d) and (e)) reveals the associated time-domain signal; the phase associated with it is essentially flat (red curve in fig. \ref{img:umbrella_v1_phases}). While the current results do not unambiguously identify the mechanism of symmetry breaking, it is a promising avenue for future exploration.

\section{Conclusion}
In conclusion, the breaking of a molecular bond is observed to induce coherent vibrational dynamics. The methyl radical, of which we study the $\nu_2$ umbrella mode, reports on these coherent dynamics in a manner governed by its symmetry; the real-space vibrational motion is mapped onto an observable energy shift as slow oscillatory motion at the difference frequencies of the vibrational progression. By constructing a model of the coherent dynamics the vibrational motion is retrieved. It is dominated by strong quantum beating due to the high degree of coherent vibrational excitation, albeit a detailed estimate of the degree of coherence is not yet possible. Mechanisms for symmetry breaking that facilitate the observation of the fundamental vibrational progression are also discussed. While the complex two-channel nature of methyl iodide photodissociation near the center of its $A$-band absorption is a challenge for the analysis in this specific experiment, the modeling presented could be applied to infer the curvature of core-excited potential energy surfaces in the future when the vibrational populations and frequencies are known.

\section*{Acknowledgements}
The experimental work is supported by the US Department of Energy (DOE) Office of Science, Basic Energy Sciences (BES) Program, Chemical Sciences, Geosciences, and Biosciences Division, under contract no. DE-AC02-05CH11231, through the Gas Phase Chemical Physics program (C.A.S., J.H.H., J.L.E., D.M.N, S.R.L.). The instrumentation was constructed using funds from the National Science Foundation through NSF MRI 1624322 and matching funds from the Lawrence Berkeley National Laboratory, the College of Chemistry, the Department of Physics, and the Vice Chancellor for Research at UC Berkeley. S.R.L. acknowledges National Science Foundation grants CHE-1951317 and CHE-2243756, which provided essential results on the conical intersection in methyl iodide and established the goal to determine coherences in bond breaking. This research also used resources of the National Energy Research Scientific Computing Center (NERSC), a Department of Energy User Facility using NERSC award BES-ERCAP0032132. C.A.S. is additionally funded by the Deutsche Forschungsgemeinschaft (DFG, German Research Foundation) - 546437684. J.H.H. acknowledges support from the Arnold and Mabel Beckman Foundation through the Arnold O. Beckman Postdoctoral Fellowship. Q.W. and H.W. utilized computing resources partially provided by the High-Performance Computing (HPC) Platform of ShanghaiTech University.

C.A.S. gratefully acknowledges enlightening discussions with Werner Fuß.

\bibliography{bibliography.bib}

\onecolumngrid
\pagebreak
\widetext
\begin{center}
\textbf{\large Supplementary material for: ``Coherent vibrational dynamics in molecular bond breaking: methyl radical umbrella mode probed by femtosecond x-ray spectroscopy"}
\end{center}

\setcounter{equation}{0}
\setcounter{section}{0}
\setcounter{figure}{0}
\setcounter{table}{0}
\setcounter{page}{1}
\makeatletter
\renewcommand{\theequation}{S\arabic{equation}}
\renewcommand{\thefigure}{S\arabic{figure}}
\renewcommand{\thetable}{S\arabic{table}}

\section{Anharmonic oscillator model for the $\nu_2$ "umbrella" mode}
\label{sec:osc}
\subsection{Hamiltonian, eigenvalues and eigenfunctions}
We use the Hamiltonian and coordinate system of Ragni \textsl{et al.} \cite{ragni2016umbrella}
\begin{equation}
    H(\vartheta) = -\frac{1}{2mr_\mathrm{eq.}^2}\frac{1}{\sin\vartheta}\frac{\partial}{\partial\vartheta}\sin\vartheta\frac{\partial}{\partial\vartheta} + V(\vartheta),
    \label{eq:hamiltonian}
\end{equation}
to represent the $\nu_2$ progression in the $\mathrm{CH_3}$ radical. There, $\vartheta$ is the hyperangle defined such that $\vartheta = \frac{\pi}{2}$ represents the planar geometry, $r_\mathrm{eq} = 1.078\,\mathrm{\AA}$ is the equilibrium C-H bond hyperradius and $m$ is total mass of the system.
\begin{figure}[h!]
    \includegraphics[width=\textwidth]{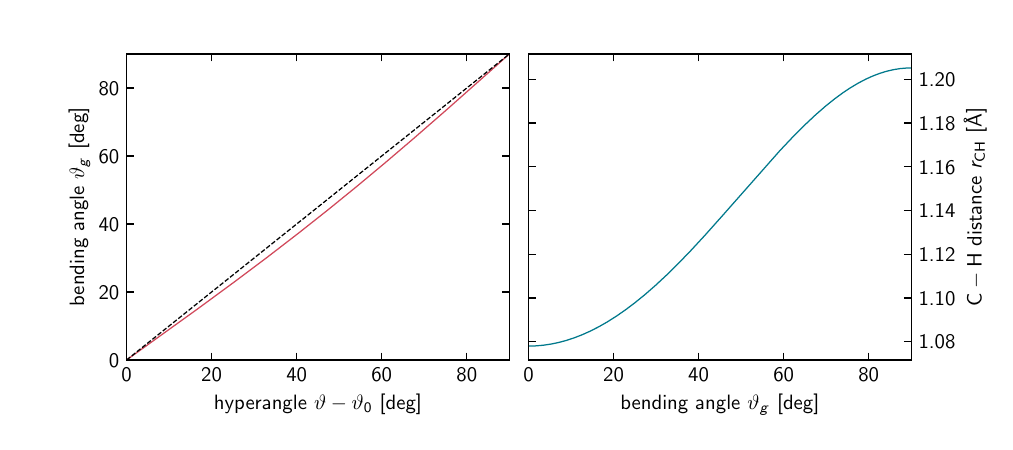}
    \caption{Relationship between hyperangle $\vartheta$ and $\nu_2$ bending angle $\vartheta_g$ (left panel). Description of the umbrella mode via the hyperangle enables factoring out the bond elongation occurring during the motion. The relationship between bending angle $\vartheta_g$ and the $\mathrm{C-H}$ bond length is shown in the right panel.}
    \label{fig:ragni_coordinates}
\end{figure}
The bending angle $\vartheta_g$ relates to the hyperangle $\vartheta$ as
\begin{equation}
\vartheta_g = \arccos \sqrt{\frac{m_\mathrm{C} \cos^2(\vartheta)}{m - 3m_\mathrm{H}\cos^2(\vartheta)}},
\end{equation}
and the bond length as a function of $\vartheta_g$ is then
\begin{equation}
r_\mathrm{CH} = \frac{r_\mathrm{eq.}}{\sqrt{1 - 3\frac{m_\mathrm{H}}{m}\cos^2(\vartheta_g)}},
\end{equation}
with $m$ being the total mass of the system and $m_\mathrm{C}$ and $m_\mathrm{H}$ being the masses of carbon and hydrogen atoms, respectively. The relationship between hyperangle $\vartheta$ and true geometrical bending angle $\vartheta_g$ as well as bond length $r_\mathrm{CH}$ is shown in fig. \ref{fig:ragni_coordinates}.

\begin{figure}[p]
    \includegraphics[width=\textwidth]{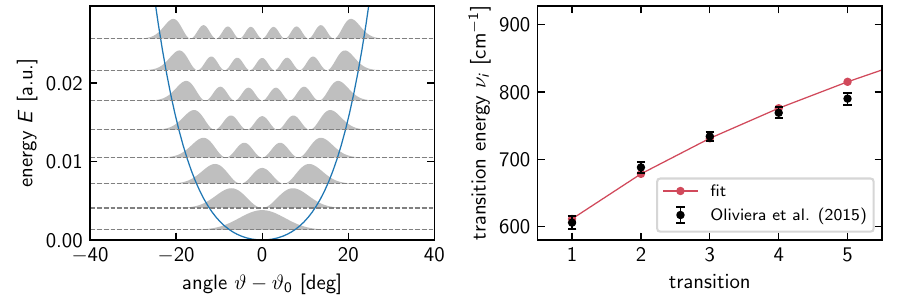}
    \caption{Potential $V(\vartheta)$ (blue, left panel), moduli of its Eigenfunctions $\left|\phi_i(\vartheta)\right|^2$ (gray, left panel) and Eigenvalues $\omega_i$ (marked with dashed black lines, left panel) for the $\mathrm{CH_3}$ radical. The transition energies $\nu_i = \omega_{i} - \omega_{i - 1}$ are compared to published values \cite{hermann1982photofragment, oliveira2015photoelectron} in the right panel.}
    \label{fig:ragni_hamiltonian}
\end{figure}

\begin{figure}[p]
    \includegraphics[width=\textwidth]{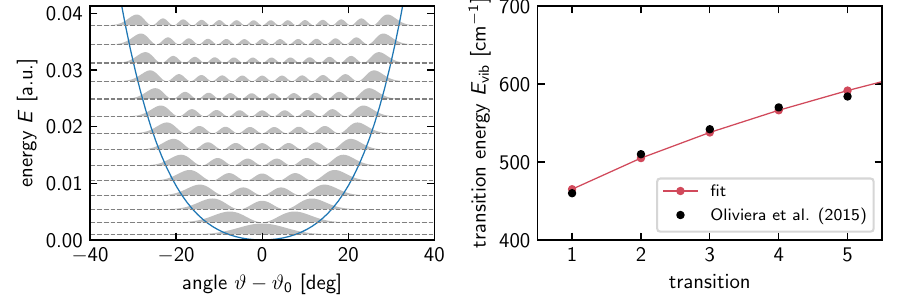}
    \caption{Potential $V(\vartheta)$ (blue, left panel), moduli of its Eigenfunctions $\left|\phi_i(\vartheta)\right|^2$ (gray, left panel) and Eigenvalues $\omega_i$ (marked with dashed black lines, left panel) for the $\mathrm{CD_3}$ radical. The transition energies $\nu_i = \omega_{i} - \omega_{i - 1}$ are compared to published values \cite{oliveira2015photoelectron} in the right panel.}
    \label{fig:ragni_hamiltonian_cd3i}
\end{figure}

The Hamiltonian eq. \ref{eq:hamiltonian} is diagonalized numerically via discretization on a regular 512 point grid spanning the interval $[\frac{1}{2}, \pi - \frac{1}{2}]$. The \texttt{eigs} function of the \texttt{scipy.sparse} module for the \texttt{python} programming language is used. The potential $V(\vartheta)$ is set up as a polynomial
\begin{equation}
    V(\vartheta) = \alpha \vartheta^2 + \beta \vartheta^4,
\end{equation}
and the parameters $\alpha$ and $\beta$ are optimized such that the transition energies $\nu_i = \omega_i - \omega_{i - 1}$ associated with eq. \ref{eq:hamiltonian} fit to published data \cite{oliveira2015photoelectron} of the umbrella progression. Figure \ref{fig:ragni_hamiltonian} shows the results of the fit and the computed transition energies for $\mathrm{CH_3}$, fig. \ref{fig:ragni_hamiltonian_cd3i} shows the same for the deuterated radical.

\subsection{Coherent superposition}
A coherent superposition can be assembled from the eigenfunctions of eq. \ref{eq:hamiltonian} as
\begin{equation}
	\psi(\vartheta, t) =
		\sum_{i=0}^{\nu_\mathrm{max}}
		\left|a_i\right|\phi_i(\vartheta)\mathrm{e}^{-i\omega_it}\mathrm{e}^{-i\varphi_i},
		\label{eq:cohr}
\end{equation}
where $a_i$ and $\varphi_i$ denote the amplitudes and phases with which the eigenfunctions $\phi_i(\vartheta)$ contribute. The expectation value of the bending angle relative to the equilibrium angle $\vartheta_0$ is then
\begin{equation}
	\langle \vartheta - \vartheta_0 \rangle = \sum_{i=0}^{\nu_\mathrm{max}}\sum_{j=0}^{\nu_\mathrm{max}} \left|a_i\right|\left|a_j\right|
	\left\langle\left.\phi_i\right.\right|\vartheta - \vartheta_0\left.\left|\phi_j\right.\right\rangle\mathrm{e}^{-i(\omega_i - \omega_j)t}\mathrm{e}^{-i(\varphi_i - \varphi_j)},
\end{equation}
which simplifies to
\begin{equation}
	\langle \vartheta - \vartheta_0 \rangle = 2 \sum_{i = 1}^{\nu}\left|a_i\right|\left|a_{i - 1}\right| \mu_{i, i - 1} \cos\left(\left(\omega_i - \omega_{i - 1}\right)t + \left(\varphi_i - \varphi_{i - 1}\right)\right)
	\label{eq:xexpct}
\end{equation}
due to symmetry, with $\mu_{i, j} = \left\langle\left.\phi_i\right.\right|\vartheta - \vartheta_0\left.\left|\phi_j\right.\right\rangle$. In the main text the equilibrium angle $\vartheta_0$ is dropped in the expressions for the position operator for notational clarity.

\section{Model function for the observed core-to-valence transition energy as a function of time}
The energy $E_\mathrm{cx}^\mathrm{exp.}(t)$ at which the C$1s\to$ SOMO transition feature appears in the experiment is the primary observable. As the total experimental signal is due to radicals generated from both the I and I${}^{*}$ channel of the dissociation, the peak positions overlap for both channels. We find that
\begin{equation}
    E_{\mathrm{exp.}}(t) \approx \sigma_\mathrm{I^*} E_\mathrm{cx}^\mathrm{I^*}(t) + (1 - \sigma_\mathrm{I^*})E_\mathrm{cx}^\mathrm{I}(t),
    \label{eq:fitfunc}
\end{equation}
where $E_\mathrm{cx}^\mathrm{I^{*}}(t)$ and $\sigma_\mathrm{I^{*}}$ denotes the momentary transition energy at time $t$ in the $\mathrm{I^{*}}$ channel and the relative population of radicals in this channel, and $E_\mathrm{cx}^\mathrm{I}(t)$ denotes the same for the $\mathrm{I}$ channel. For each channel the transition energy $E_\mathrm{cx}(t)$ as a function of time is obtained by inserting eq. \ref{eq:xexpct} into a function $E_\mathrm{ce}(\vartheta)$ that describes how the core-excited state energy varies as a function of bending angle $\vartheta$, with the appropriate amplitudes $a_i$ and phases $\varphi_i$ set, and finally convolved with the temporal instrument response function for which we use a Gaussian with an FWHM of $36\,\mathrm{fs}$ (see sec. \ref{sec:apparatus}).

\subsection{Parameters of the model}
We take the amplitudes $|a_i|^2 = p_i$ for each channel from the published populations $p_i$ determined in the photofragmentation studies also cited in the main text (refs. \cite{eppink1999energy, aguirre2005photoionization, cheng2011vibrationally, li2005high} for $\mathrm{CH_3}$ and ref. \cite{li2006state} for $\mathrm{CD_3}$). From there we also take the channel branching ratios and use $\sigma_\mathrm{I^{*}} = 0.75$ throughout for $\mathrm{CH_3I}$ and $\sigma_\mathrm{I^{*}} = 0.81$ for $\mathrm{CD_3I}$. The frequencies $\omega_i$ and matrix elements $\mu_{i, j}$ are determined from the oscillator model (sec. \ref{sec:osc}). The free parameters of the model, which are ultimately determined in the fit, are the phases $\varphi_i$, an overall energy shift $\Delta E$, a global time shift $\Delta t$ and a phenomenological scaling factor $\kappa$ for eq. \ref{eq:xexpct}. All parameters from the fit for all cited population data as well as both channels are listed in tables \ref{tab:phases_istar}, \ref{tab:phases_i} and \ref{tab:fitparams}.

\section{Fitting procedure}
\begin{figure}
    \includegraphics[width=\textwidth]{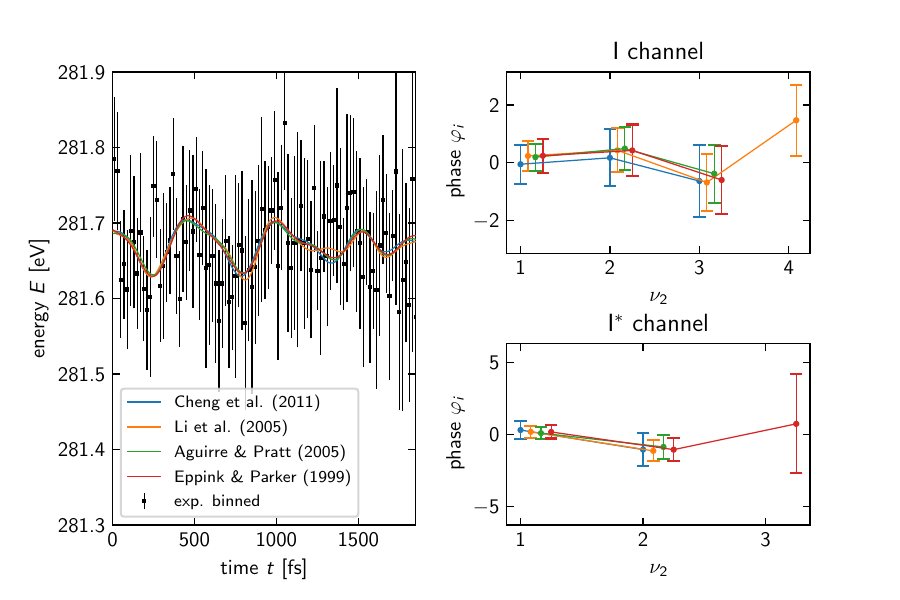}
    \caption{A comparison of the model fit results  in $\mathrm{CH_3}$ for all cited vibrational distributions \cite{cheng2011vibrationally, li2005high, aguirre2005photoionization, eppink1999energy} and the raw experimental data (points, left panel). Agreement between the individual fits is excellent. The right panel shows the phases reported by the fit, with the respective uncertainties. A slight horizontal shift is introduced per dataset for visual clarity. The numerical data presented in the right panels is listed in tables \ref{tab:phases_i} and \ref{tab:phases_istar}. We enforce $\varphi_0=0$ in the fit.}
    \label{fig:fits_comparison}
\end{figure}

We fit eq. \ref{eq:fitfunc} to the experimentally determined energy shift of the radical peak. The convolution with the fixed-width instrument response function results in a smoothing of the fit function and renders the fit insensitive to fast variations of the experimental data that are present e.g. due to noise, or those that we tentatively attribute to the symmetry breaking (see main text). We bin the experimental data (two consecutive points) as we find it to improve convergence. As initial values, all phases $\varphi_i$ are set to zero. The fit is restricted to the experimental data where $\tau > 150\,\mathrm{fs}$ in order to ensure that an energetic shift of the radical peak due to the initial vicinity of the iodine atom has no influence. A comparison between the fits performed using the different population data \cite{cheng2011vibrationally, li2005high, aguirre2005photoionization, eppink1999energy} is given in fig. \ref{fig:fits_comparison}.

\begin{table}
    \centering
    \caption{Phases $\varphi_i$ (in radians) obtained for the I channel in $\mathrm{CH_3}$.}
    \label{tab:phases_i}
\begin{tabular}{l|c|c|c|c}

 $\nu_2$ population data & 1                             & 2                               & 3                              & 4                              \\
 \hline
 \hline
 Cheng et al. (2011)     & $(-0.1 \pm 0.7)$ & $(0.2 \pm 1.0)$ & $(-0.6 \pm 1.2)$ &                 \\
 Li et al. (2005)        & $(0.2 \pm 0.5)$  & $(0.4 \pm 0.8)$ & $(-0.7 \pm 1.0)$ & $(1.5 \pm 1.2)$ \\
 Aguirre \& Pratt (2005) & $(0.2 \pm 0.5)$  & $(0.5 \pm 0.7)$ & $(-0.4 \pm 1.0)$ &                 \\
 Eppink \& Parker (1999) & $(0.2 \pm 0.6)$  & $(0.4 \pm 0.9)$ & $(-0.6 \pm 1.2)$ &                 \\
\end{tabular}
\end{table}

\begin{table}
    \centering
    \caption{Phases $\varphi_i$ (in radians) obtained for the I${}^{*}$ channel in $\mathrm{CH_3}$.}
    \begin{tabular}{l|c|c|c}
 $\nu_2$ population data & 1                 & 2                 & 3                 \\
 \hline
 \hline
 Cheng et al. (2011)     & $(0.3 \pm 0.6)$ & $(-1.1 \pm 1.1)$ &                 \\
 Li et al. (2005)        & $(0.2 \pm 0.4)$ & $(-1.1 \pm 0.7)$ &                 \\
 Aguirre \& Pratt (2005) & $(0.1 \pm 0.4)$ & $(-0.9 \pm 0.8)$ &                 \\
 Eppink \& Parker (1999) & $(0.2 \pm 0.5)$ & $(-1.1 \pm 0.8)$ & $(0.7 \pm 3.4)$ \\
\end{tabular}
    \label{tab:phases_istar}
\end{table}

\begin{table}
    \centering
    \caption{Residual fit parameters for the individual population data.}
    \begin{tabular}{l|c|c|c}
    \hline
                            & $E_0$ [eV]          & $E_1$ [$\mathrm{eV / rad^2}$] & $t_0$ [fs]        \\
    \hline
    \hline
 Cheng et al. (2011)     & $(281.71 \pm 0.01)$ & $(-9.0 \pm 2.2)$              & $(71.1 \pm 11.6)$ \\
 Li et al. (2005)        & $(281.73 \pm 0.01)$ & $(-12.7 \pm 2.8)$             & $(74.8 \pm 7.7)$  \\
 Aguirre \& Pratt (2005) & $(281.73 \pm 0.02)$ & $(-14.1 \pm 3.7)$             & $(74.3 \pm 7.7)$  \\
 Eppink \& Parker (1999) & $(281.72 \pm 0.01)$ & $(-10.2 \pm 2.4)$             & $(74.5 \pm 9.5)$  \\
    \hline
    \end{tabular}
    \label{tab:fitparams}
\end{table}

\section{Other photoproducts}
While the resonant dissociative transition from the electronic ground state into the $A$-band strongly dominates in the experiment, the influence of other possible pathways and photoproducts must be considered. The most likely source of spectral contamination in the experiment is the production of methyl iodide cations produced via multiphoton ionization due to the strong pump field. The $\tilde A$ state of the $\mathrm{CH_3I^{+}}$ cation can be accessed via absorption of 3 photons of the $266\,\mathrm{nm}$ pump (cf. \cite{powis1983unimolecular, walter1988molecular}), and from there the vibrationally excited $\mathrm{CH_3I^{+}}$ can undergo dissociation into either $\mathrm{CH_3}$ radicals and $\mathrm{I^+}$ or $\mathrm{CH_3^{+}}$ radicals and $\mathrm{I}$. The branching ratio for these competing channels is determined by the excess energy available \cite{olney1998quantitative}, and at small excess energies, provided by absorption of $3\times 266\,\mathrm{nm}$ photons, the pathway producing $\mathrm{CH_3^{+}}$ and $\mathrm{I}$ is strongly favored over that producing $\mathrm{CH_3}$ radicals and $\mathrm{I^+}$ by a factor of $\sim 10-15$. As such, even if, say $10-20\%$ of the transient signal is due to methyl iodide cations (i.e. a very generous estimate), the $\mathrm{CH_3}$ yield in this channel will be on the order of a single-digit percentage and therefore negligible.

\begin{figure}
    \includegraphics[width=\textwidth]{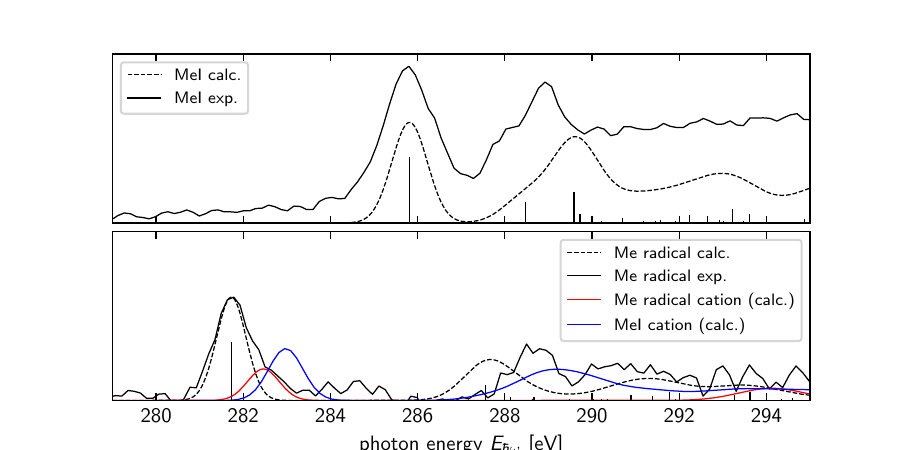}
    \caption{Calculated x-ray absorption for methyl iodide (MeI), its cation, the methyl radical (Me), and the methyl radical cation compared to the experiment. Both cations ($\mathrm{CH_3I^{+}}$ and $\mathrm{CH_3^{+}}$) appear at different energies than the methyl radical. Furthermore, the high-energy shoulder of the experimentally observed radical peak may be attributable in part to the presence of $\mathrm{CH_3^{+}}$ cations.}
    \label{fig:stobe}
\end{figure}

Additionally, both $\mathrm{CH_3^{+}}$ cations and $\mathrm{CH_3I^{+}}$ cations are spectroscopically distinct from the $\mathrm{CH_3}$ radical in their x-ray absorption signature as confirmed via calculations using the \texttt{StoBe-deMon} software package \cite{stobe}, as shown in fig. \ref{fig:stobe}. We choose the IGLO-III basis set for the carbon atoms, IGLO-II for the iodine atom and STO-311 Gaussian basis for hydrogen.  The carbon atom's basis is augmented by the means offered in the \texttt{StoBe-deMon} program. A global energy scale for these calculations is established by correcting the calculated transition energies by the difference $\Delta E = I_\mathrm{p}^{\mathrm{hole}} - I_\mathrm{p}^{\mathrm{TP}}$ of ionization potential $I_\mathrm{p}^\mathrm{TP}$ determined via a transition-potential type calculation and as determined from a calculation of the pure hole-state and the ground state ($I_\mathrm{p}^\mathrm{hole} = E_{\mathrm{C}1s - 1} - E_\mathrm{gs}$). The input and output files as well as the processing scripts for these calculations are included in the data repository. An energy shift of $2.6\,\mathrm{eV}$ towards lower energies is applied to the calculated ground state spectrum, which aligns the $\mathrm{C}1s\to\mathrm{HOMO}$ transition to the experimental value. For the radical and cation species we apply an overall energy shift of $4.55\,\mathrm{eV}$ towards lower energies, which aligns the $\mathrm{C}1s\to\mathrm{SOMO}$ transition in the radical to the experimental value.

\section{Extended experimental data}
\begin{figure}
    \includegraphics[width=\textwidth]{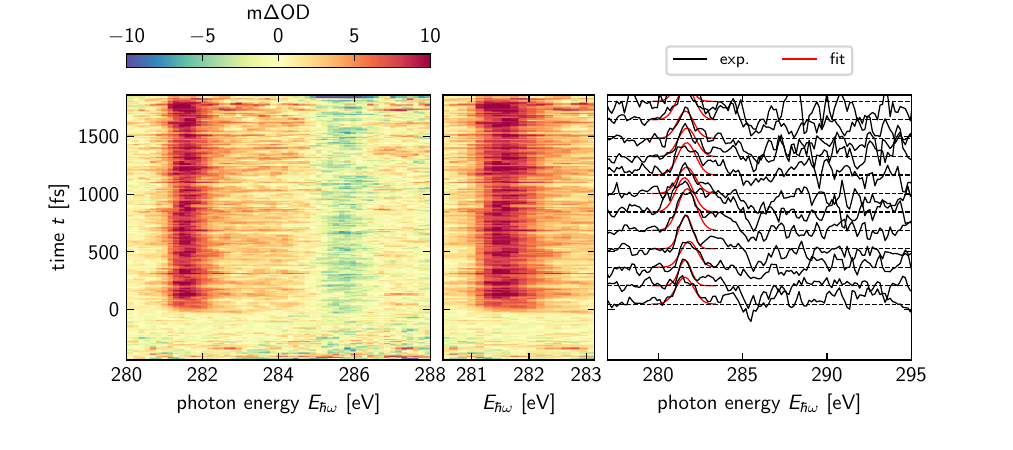}
    \caption{The averaged experimental transient absorption map for $\mathrm{CH_3I}$. The left panel shows the entire near-edge region with both the emerging radical peak as well as the ground state depletion visible. The center panel is a zoom-in on the radical peak showing a small but discernible time-dependent motion. In the right panel we show a subset of time slices that illustrate how the energetic position of the radical peak is determined by fitting a Gaussian function to it.}
    \label{img:trace_ch3i}
\end{figure}

\begin{figure}
    \includegraphics[width=\textwidth]{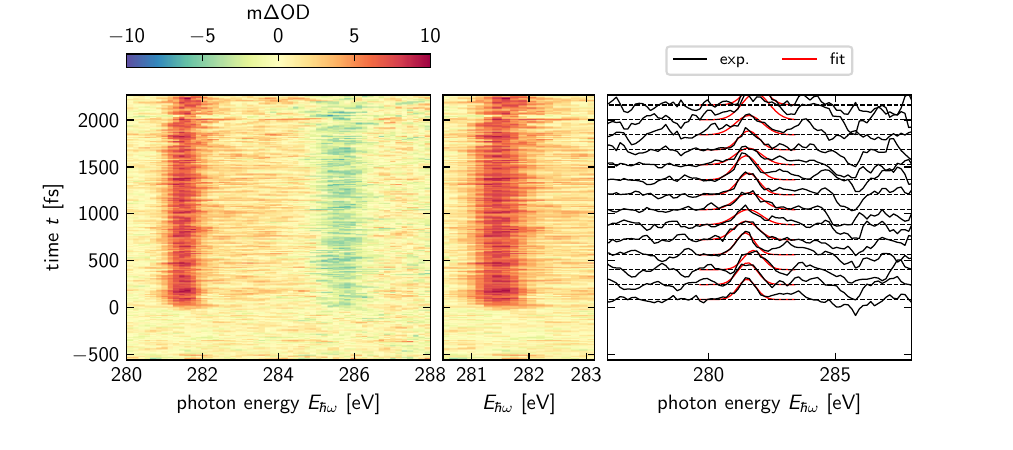}
    \caption{The averaged experimental transient absorption map for $\mathrm{CD_3I}$. The left panel shows the entire near-edge region with both the emerging radical peak as well as the ground state depletion visible. The center panel is a zoom-in on the radical peak showing a small but discernible time-dependent motion. In the right panel we show a subset of time slices that illustrate how the energetic position of the radical peak is determined by fitting a Gaussian function to it.}
    \label{img:trace_cd3i}
\end{figure}

The averaged transient absorption traces are shown in figures \ref{img:trace_ch3i} and \ref{img:trace_cd3i}. The raw dataset is available online.

\section{Characterization of the apparatus}
\label{sec:apparatus}
\begin{figure}
    \includegraphics[width=\textwidth]{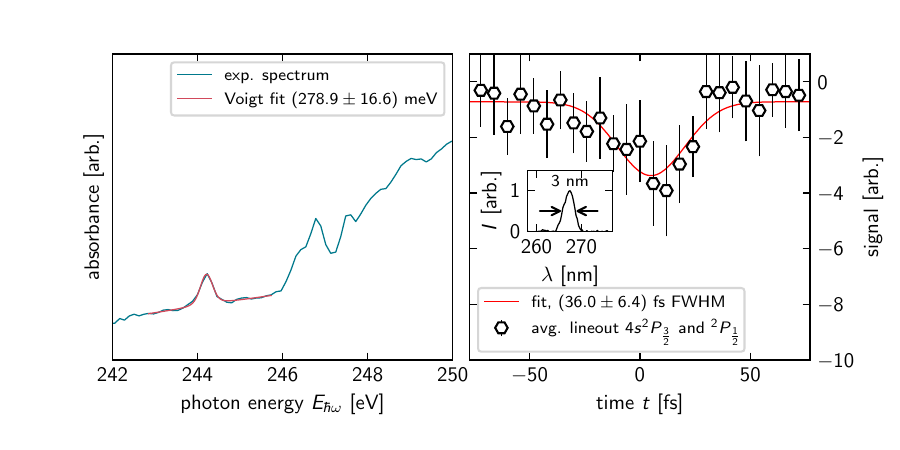}
    \caption{Static absorption of argon gas at its $L_{2, 3}$ edge (left panel) and SXR/UV cross correlation measured via ponderomotive supppression of the $4s{}^{2}P_\frac{3}{2}$ and ${}^{2}P_\frac{1}{2}$ lines (right panel) as measured in the apparatus. The UV spectrum is shown in the inset.}
    \label{img:setup_characterization}
\end{figure}

We characterize the spectral and temporal resolution of the apparatus using the absorption of argon gas around the Ar$L_{2, 3}$ edge ($240-250\,\mathrm{eV}$, see fig. \ref{img:setup_characterization}). The linewidths of the $L_{2,3}$ absorption spectra are known from literature \cite{king1977investigation}, and assuming a gaussian spectral response the absorption features should appear as Voigt profiles. A Voigt profile with a linear baseline is fit to the $4s{}^{2}P_\frac{3}{2}$ absorption line, keeping the natural linewidth of the transition fixed to the literature value of $121\,\mathrm{meV}$ and the apparatus is determined to have an energy resolution of $(278.9\pm16.6)\,\mathrm{meV}$ in this specific measurement. This precise value does change day-to-day depending on alignment, but around $300\,\mathrm{meV}$ is typical. In order to determine the temporal response function of the apparatus we use the UV pump pulse to ponderomotively suppress the Ar$4s{}^{2}P_\frac{3}{2}$ and Ar${}^{2}P_\frac{1}{2}$ absorption. As a function of delay time between UV and soft x-ray pulse this records their cross-correlation, which has a FWHM of $(36\pm 6)\,\mathrm{fs}$. Given that the spectral width of the probe pulse is $3\,\mathrm{nm}$ (centered at $267\,\mathrm{nm}$) this indicates that the UV pulses are nearly bandwidth limited and the soft x-ray pulses are much shorter.

\section{Core-excited state calculation}
\begin{figure}
    \centering
    \includegraphics{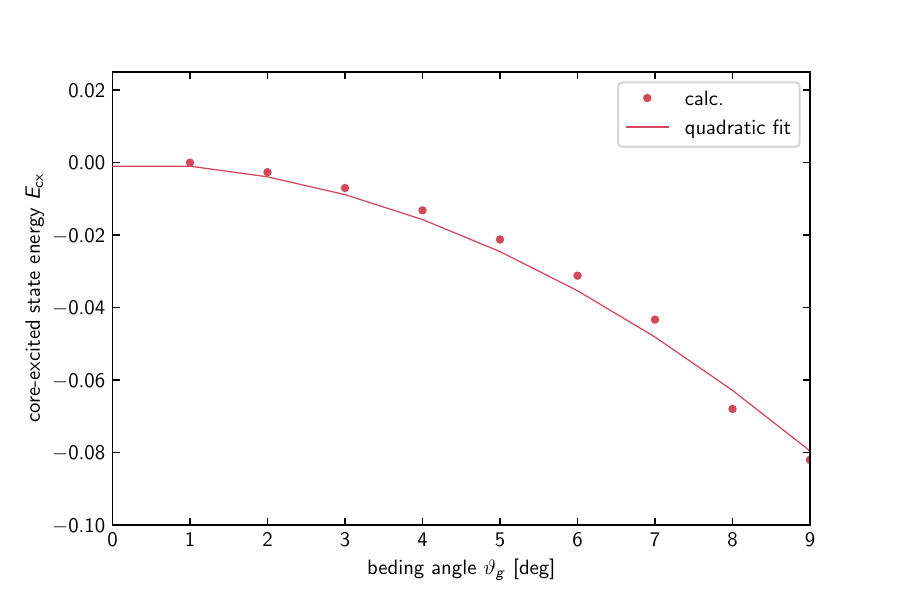}
    \caption{Core-excited state energy as a function of bending angle as calculated (dots) and quadratic fit (solid line).}
    \label{img:compare_surfaces}
\end{figure}
X-ray absorption spectra (XAS) were computed using the MS-RASPT2 \cite{finley1998multi, sauri2010multiconfigurational} and restricted active space state interaction (RASSI) methods \cite{malmqvust1989casscf} as implemented in OpenMolcas \cite{openmolcas}, employing the ANO-RCC-VTZP basis set throughout. The carbon 1s orbital is placed in RAS1 and frozen from the PBE0 level. State-averaged RASSCF orbitals were optimized by including 8 valence orbitals and 7 electrons in the RAS2 active space. 10 roots were computed for the valence excited states, while 30 core-excited states were calculated for the XAS simulations using the "highly-excited-state (HEXS)" scheme. Figure \ref{img:compare_surfaces} shows the core-excited state energy calculated this way as a function of the bending angle.

\section{The effect of deuteration}
\begin{figure}
    \centering
    \includegraphics{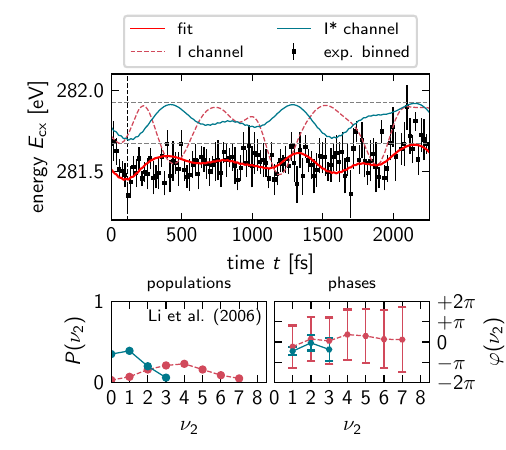}
    \caption{Application of the two-channel model to the photodissociation of $\mathrm{CD_3I}$ using the populationd data of \cite{li2006state}.}
    \label{img:model_cd3i}
\end{figure}

Deuteration of the methyl group in $\mathrm{CD_3I}$ changes the dynamics of the photodissociation process significantly. Vibrational excitation to much higher levels than for $\mathrm{CH_3I}$ has been reported  for $266\,\mathrm{nm}$ photodissociation \cite{li2006state}. We show the application of our two-channel model to $\mathrm{CD_3I}$ in fig. \ref{img:model_cd3i}, finding reasonable agreement. However, the significantly higher level of vibrational excitation may facilitate intramolecular vibrational energy transfer from the $\nu_2$ mode in the $\mathrm{I}$ channel into e.g. the symmetric stretch or deformation mode. Estimating the total amount of vibrational energy in both channels we find $590.4\,\mathrm{cm^{-1}}$ ($800-1100\,\mathrm{cm^{-1}}$) and $2091.7\,\mathrm{cm^{-1}}$ ($350-400\,\mathrm{cm^{-1}}$) for $\mathrm{CD_3I}$ ($\mathrm{CH_3I}$) for the $\mathrm{I^*}$ and $\mathrm{I}$ channel, respectively. For the deuterated radical the total energy in the $\mathrm{I}$ channel is sufficient to excite the $\nu_4'$ d-deform mode and is very close to the excitation energy of the $\mathrm{CH}$ stretch vibration. In $\mathrm{CH_3I}$ the total vibrational energy in either channel is not sufficient to excite any other vibrational mode (cf. \cite{eppink1998methyl}). To this end it is not clear whether the $\mathrm{CD_3I}$ results are affected by intramolecular energy redistribution and they must therefore be interpreted carefully.

\end{document}